\documentclass[pdftex]{article}
\usepackage{icrctc07}

\title{Monte Carlo Simulation for the MAGIC-II System}
\shorttitle{MAGIC-II Monte Carlo simulation}
\authors{E. Carmona$^{1}$, P. Majumdar$^{1}$, A. Moralejo$^{2}$,
V. Vitale$^3$, D. Sobczynska$^4$, M. Haffke$^5$, C. Bigongiari$^6$,
N. Otte$^{1}$, G. Cabras$^3$, M. De~Maria$^3$, F. De~Sabata$^3$ for the MAGIC Collaboration}
\shortauthors{}
\afiliations{$^1$Max-Planck-Institut f\"ur Physik\\ $^2$Institut de
F\'{\i}sica d'Altes Energies\\ $^3$Dipartimento di Fisica dell'Universit\`a
di Udine and INFN sez. di Trieste\\ $^4$Division of Experimental
Physics, University of Lodz\\ $^5$Universit\"at Dortmund\\ $^6$IFIC -
Instituto de F\'{\i}sica Corpuscular} 

\email{carmona@mppmu.mpg.de, pratik@mppmu.mpg.de}

\abstract{Within the year 2007, MAGIC will be upgraded to a two telescope
system at La Palma. 
Its main goal is to improve the
sensitivity in the stereoscopic/coincident operational mode. At the
same time it will lower the analysis threshold of the currently
running single MAGIC telescope. Results from the Monte Carlo
simulations of this system will be discussed. A comparison of the two
telescope system with the performance of one single telescope will be
shown in terms of sensitivity, angular resolution and energy
resolution.}

\begin{document}
\maketitle

\section{Introduction}

The MAGIC telescope, currently the largest (17~m diameter mirror)
single dish imaging atmospheric Cherenkov telescope, has been in
scientific operation since summer 2004.  A second 17~m telescope
equipped with advanced photodetectors and ultra-fast readout is under
construction and is expected to be ready in 2007~\cite{magic2}. MAGIC-II,
the two telescope system, is designed to lower the energy threshold
and simultaneously achieve a higher sensitivity in
stereoscopic mode.

In order to determine the optimal baseline distance between the first
and the second telescope and estimate the performance of the system,
detailed Monte Carlo simulations have been carried
out~\cite{abelardo}. The Monte Carlo simulation of the MAGIC-II system
is divided into three stages. The {\em CORSIKA}~\cite{corsika} program
simulates the air showers initiated by either high energy gammas or
hadrons.  In this simulation we have used the CORSIKA version 6.019,
the EGS4 code for electromagnetic shower generation and VENUS and
GHEISHA for high and low energy hadronic interactions
respectively. New atmospheric models have been introduced on the basis of
studies of total mass density as a function of the height. The second
stage of the simulation, {\em Reflector } program, accounts for the
Cherenkov light absorption and scattering in the atmosphere and then
performs the reflection of the surviving photons on the mirror dish. Finally,
the {\em Camera} program simulates the behaviour of the MAGIC
photomultipliers, trigger system and data acquisition
electronics. Pulse shapes, noise levels and gain fluctuations obtained
from the MAGIC data have been implemented in the simulation.

For the present study a total of 1.14 $\times$ 10$^8$ protons between
30 GeV and 30 TeV have been produced, as well as and 2.0 $\times$
10$^6$ gammas between 10 GeV and 20 TeV. The energy distribution of
primary gamma rays is a pure power law with a spectral index of -2.6, whereas
charged primaries follow the power law of -2.78. The telescope
pointing direction is 20$^\circ$ in zenith, with the directions of
protons scattered isotropically within a $5^\circ$ semi-aperture cone
around the telescope axis. Maximum impact parameters of 350 and 450~m
have been simulated for gammas and hadrons respectively.

\section{Analysis of stereo events}

The two telescopes in the MAGIC-II system can be independently
operated by observing two different sources or sky regions. However,
the best performance of the system is achieved with the simultaneous
observation of air showers by the two telescopes. The stereoscopic
observation mode allows a more precise reconstruction of the shower
parameters as well as a stronger suppression of the hadronic showers and
other background events.

\begin{figure}
\begin{center}
\includegraphics [width=\columnwidth]{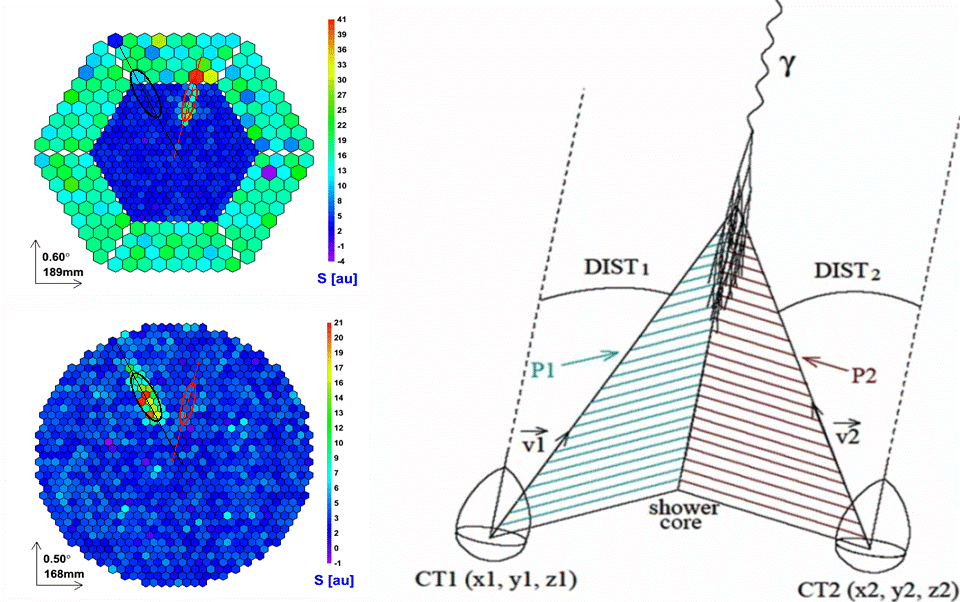}
\end{center}
\caption{Stereoscopic principle. The incoming direction of the shower and core
  distance are obtained by combining the information of the images on
  both telescopes.}
\label{fig: principle}
\end{figure}

The analysis of stereoscopic events is performed by individually analyzing the
images from the two telescopes. A set of parameters (Hillas
parameters~\cite{hillas1985}) is obtained from each image and they are
combined to obtain the shower parameters. Only showers triggering both
telescopes are considered under the stereo analysis. The images are
combined following the first algorithm in~\cite{hofmann-1999-122}. The
intersection point of the two major axis of the ellipses recorded in
the telescope cameras, provides the location of the source of a
particular shower (figure~\ref{fig: principle}). The $\theta^2$
parameter is defined as the square of the angular distance from the real source
image in the camera and the reconstructed one for each event. The
location of the shower core on ground is obtained by intersecting the
image axes from the telescope positions on the ground. In addition,
the height of the shower maximum ($H_{max}$) can also be
obtained. Having only two telescopes, the quality of the
reconstruction of these parameters depends on the amplitude ($Size$)
of the recorded images and the angle between the axes.

Image analysis is based on the analysis of the second moments of the
images recorded by the telescopes cameras. Among the parameters
obtained, {\em Width} and {\em Length} (second moments of the light
distribution along the major and minor axes of the image) are the more
useful ones for signal (gammas) / background (hadrons) separation. {\em
Width} and {\em Length} depend on the distance from the shower core to
the telescope, hence, a correct estimation of the impact parameter is
required to properly evaluate these parameters. With a single
telescope, the observer can not easily resolve the ambiguity between a
close by, low energy shower and a distant, high energy one. With a
second telescope, in most cases the ambiguity disappears because of
the stereoscopic vision of the showers.

\begin{figure}
\begin{center}
\includegraphics [width=\columnwidth, height=0.95\columnwidth]{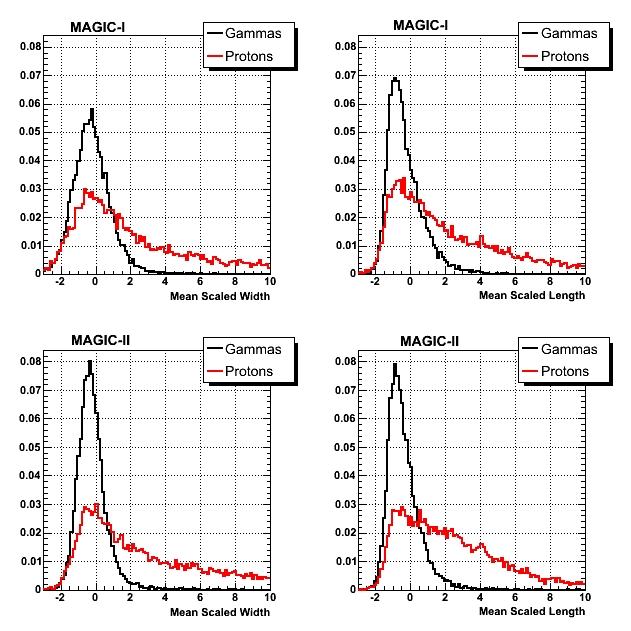}
\end{center}
\caption{Comparison of the scaled $Width$ (left) and $Length$ (right)
  parameters between MAGIC-I and MAGIC-II.}
\label{fig: hillas}
\end{figure}

In order to combine the parameters from both images, we compute the
{\em Mean Scaled Width}(MSW) and {\em Length}(MSL) parameters. These
new parameters are obtained by subtracting the mean and dividing by
the RMS of the parameter distribution (as a function of size) for
Monte Carlo gammas. The new distributions have a mean value of 0 and
RMS of 1 for gamma showers and are broader for proton showers.  The
distribution of these parameters for gamma and proton showers are
shown in figure~\ref{fig: hillas}. A comparison between the single
telescope case (MAGIC-I) and two telescopes case (MAGIC-II) is also
shown. A higher gamma/hadron separation is achieved when combining the
information from the two telescopes, compared with the single
telescope case.

In our analysis, the {\em Random Forest}~\cite{RF} ({\em RF}) technique is used for
gamma/hadron separation. {\em RF} is a multidimensional classification
tool that, in this case, it is used to determine an average
probability of an event to be a hadron induced shower. A set of MC
gamma ray events and MC proton events is used to train the {\em
RF}. After training, the test samples can be classified by {\em RF},
providing a parameter (called $Hadronness$) distributed between $0$
and $1$. Values closer to $0$ mean that the event is more gamma-like
and values closer to $1$ mean that the event is more hadron-like.

For this study, the parameters that have been used in Random Forest
are: average amplitude ($Size$), average number of islands
($N_{island}$)~\footnote{an island is defined as any cluster 
of 2 or more pixels surviving the image cleaning.}, core distance, $H_{max}$, $MSW$ and
$MSL$. This set of parameters, however, should not be considered as
optimal and some improvement could be expected with an optimized
parameter selection.

In the analysis, gamma/hadron separation is based on the
$Hadronness$ and $\theta^2$ parameter is used to extract the signal events.


Finally, energy reconstruction is based on lookup-tables where Monte
Carlo energy of gamma showers is tabulated as a function of shower
impact parameter, {\em $H_{max}$} and $Size$. For each telescope, a
reconstructed energy is obtained by interpolation. The shower energy
is obtained as the average of the values from both telescopes.

\section{Simulation results}

Monte Carlo simulations show that the sensitivity does not vary
dramatically with the distance between telescopes, the optimal value
being around 90~m~\cite{abelardo}.  The sensitivity is defined as
``integral flux resulting in gamma excess events, in 50 hours of
observation, equals to 5 times the standard deviation of the
background".  The effective area at trigger level of the MAGIC-II
system is smaller than that of any of the two telescopes because of
the coincidence requirement. However, at energies above 300~GeV
matches the effective area of MAGIC-I which is slightly lower than
that of the new second telescope because this is equipped with higher
quantum efficiency photomultipliers. The coincidence requirement
provides a higher background rejection.  As a result a better
sensitivity below 100 GeV and also a reduction of the analysis
threshold is achieved.

\begin{figure}
\begin{center}
\includegraphics [width=\columnwidth]{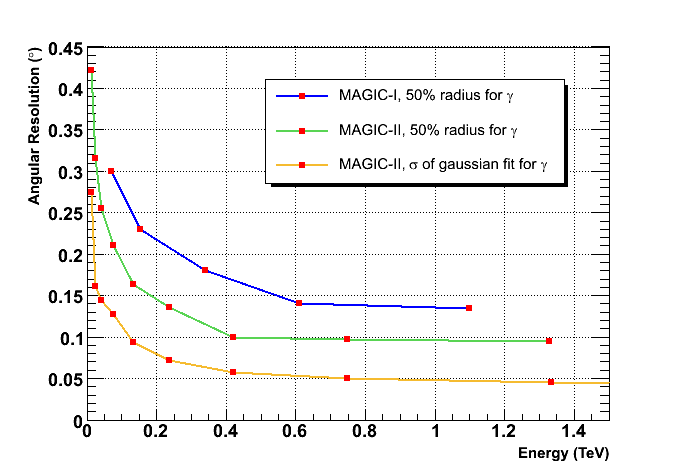}
\end{center}
\caption{Comparison of the angular resolution between MAGIC-I (1 telescope)
  and MAGIC-II (two telescopes). Sigma of the fit to a 2-dimensional gaussian is also shown for MAGIC-II.}
\label{fig: angle}
\end{figure}

For a single telescope, the angular resolution is estimated using a
modified parametrisation of the so called {\it $DISP$}
method~\cite{domingo}. The discrimination of the shower head and tail
relies on the shape of the image (asymmetry along the major
axis). This often results in a wrong head-tail assignment that degrades
the angular resolution. With two telescopes, this drawback is easily
overcome since the source direction is obtained as the intersection of
major axes of the images in the camera. The angular resolution, here
defined as the angle within which 50\% of the reconstructed gammas
from a point source would be contained, as a function of gamma ray
energy is shown in figure~\ref{fig: angle}. The improvement in angular
resolution from 1 to 2 telescopes is clearly seen.

The stereoscopic analysis also results in a better energy
reconstruction due to better reconstruction of the shower axis and
also a double sampling of the light pool. The energy resolution for
gammas as a function of primary energy is shown in figure~\ref{fig:
energy}. For comparison, the energy resolution of MAGIC-I is also
shown. An energy resolution for gammas better than 20\% is achieved
above 50 GeV.

\begin{figure}
\begin{center}
\includegraphics [width=\columnwidth]{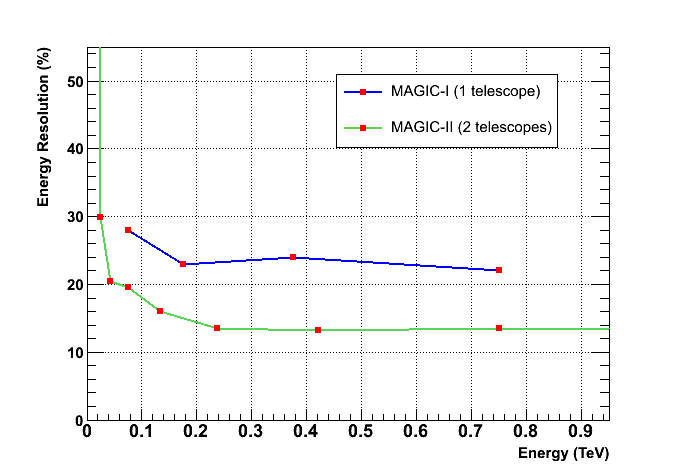}
\end{center}
\caption{Energy resolution of the MAGIC-II (two telescopes) system
  compared with MAGIC-I.}
\label{fig: energy}
\end{figure}

The sensitivity estimate for MAGIC-II is shown in figure~\ref{fig:
sensitivity}. The flux sensitivity of the 2-telescope system is
between 2 and 3 times better than that of a single telescope (MAGIC-I)
and it is significantly improved below 100 GeV. The MAGIC-II system
can achieve a sensitivity of 1\% Crab in 50 hours above 150~GeV.
  
\begin{figure}
\begin{center}
\includegraphics [width=\columnwidth]{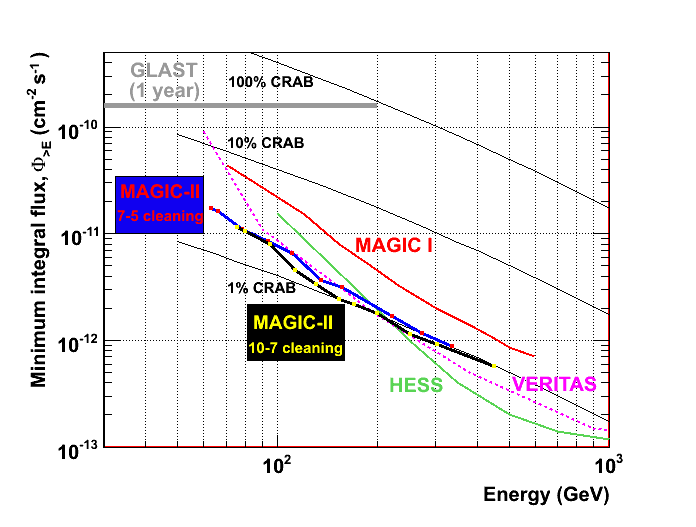}
\end{center}
\caption{MAGIC-II system sensitivity compared with the Crab flux and
  other existing experiments (MAGIC-I, HESS and VERITAS).}
\label{fig: sensitivity}
\end{figure}

\section{Conclusions}

The MAGIC-II system of 2 telescopes will perform observations in
stereoscopic mode, allowing a more precise reconstruction of the
showers and a significant reduction of backgrounds below 100 GeV. This
will make possible an improved angular and energy resolution as well
as a reduction of the analysis threshold. All together it will
increase the current sensitivity of the instrument by a factor between
2 and 3 at different energies.

\section{Acknowledgements}

 The authors thank other collaborators of MAGIC for valuable discussions and 
gratefully acknowledge the support of MPG and BMBF
(Germany), the INFN (Italy) and the Spanish CICYT for this work.
This work was also supported by ETH Research Grant TH34/043 and the Polish
MNil Grant 1P03D01028

\bibliography{icrc0588.bib}
\bibliographystyle{unsrt}

\end{document}